\documentclass[nofootinbib,superscriptaddress,twocolumn,showpacs,preprintnumbers,amsmath,amssymb]{revtex4-1}

\usepackage{graphicx}
\usepackage{grffile}
\usepackage{slashed}
\usepackage{bbm}
\usepackage{footmisc}
\usepackage{multirow}

\usepackage{hyperref}
\usepackage{amssymb}
\usepackage{amsmath}
\usepackage{color}
\usepackage{xspace}
\usepackage{caption}
\usepackage{subcaption}
\usepackage{natbib}

\usepackage{datetime}
\usepackage{color,fancybox}
\newcommand{\GeV}{\ensuremath{\,\mathrm{GeV}}\xspace}

\newcommand{\fb}{\ensuremath{\,\mathrm{fb}}\xspace}
\newcommand{\MS}{\ensuremath{\,\mathrm{MS}}\xspace}

\newcommand{\LO}{\ensuremath{\,\mathrm{\text{LO}}\xspace}}
\newcommand{\NLO}{\ensuremath{\,\mathrm{\text{NLO}}\xspace}}

\newcommand{\order}[1]{\mathcal{O}\!\left(#1\right)}

\newcommand{\bea}{\begin{eqnarray}}
\newcommand{\eea}{\end{eqnarray}}

\newcommand{\bib}[1]{Ref.~\cite{#1}}
\newcommand{\fig}[1]{Fig.~\ref{#1}}
\newcommand{\tab}[1]{Table~\ref{#1}}
\newcommand{\sect}[1]{Section~\ref{#1}}

\begin{document}
\title{$Z\gamma$ production in association with two jets at next-to-leading order QCD}

\preprint{FTUV-14-0729\;\; IFC/14-52\;\; KA-TP-20-2014\;\;LPN14--098\;\;SFB/CPP-14-58}

\author{Francisco~Campanario}
\email{francisco.campanario@ific.uv.es}
\affiliation{Theory Division, IFIC, University of Valencia-CSIC, E-46980
  Paterna, Valencia, Spain}
\author{Matthias~Kerner}
\email{matthias.kerner@kit.edu}
\affiliation{Institute for Theoretical Physics, KIT, 76128 Karlsruhe, Germany}
\author{Le~Duc~Ninh}
\email{duc.le@kit.edu}
\affiliation{Institute for Theoretical Physics, KIT, 76128 Karlsruhe, Germany}
\author{Dieter~Zeppenfeld}
\email{dieter.zeppenfeld@kit.edu}
\affiliation{Institute for Theoretical Physics, KIT, 76128 Karlsruhe, Germany}

\begin{abstract}
Next-to-leading order QCD corrections to the QCD-induced 
$pp \rightarrow l^+l^- \gamma j j +X$ and $pp \rightarrow \bar{\nu}_l \nu_l \gamma jj+X$ 
processes are presented. The latter is used to find an optimal cut to reduce the contribution of 
radiative photon emission off the charged leptons in the first channel. 
As expected, the scale uncertainties are significantly reduced at NLO and the QCD corrections are 
phase space dependent and important for precise measurements at the LHC. 
\end{abstract}

 % perturb. calc., high E, Z boson

%\vspace*{\fill} {\bf \Large \today}

\maketitle

\section{Introduction}
\label{sec:intro}
The production of a prompt photon in association with two charged leptons and two jets at the LHC is an
attractive mechanism to study weak boson scattering, namely $W^+ W^- \to \gamma V$ with
$V = Z/\gamma^*$.
% and the initial $W$ bosons being radiated from the quarks.
It is also
relevant to the study of anomalous gauge boson couplings, which may provide hints of new physics beyond
the Standard Model (SM).

At leading order (LO), the process 
$pp \rightarrow jj \gamma l^+l^- +X $ is classified into two mechanisms: 
the electroweak-induced mechanism of
order $\order{\alpha^5}$, which is sensitive to $W^+ W^- \to \gamma V$ scattering
and the QCD-induced channel of order $\order{\alpha_s^2 \alpha^3}$, which can be considered
as a background. The EW contributions can be further classified into $t$-channel
vector-boson fusion contributions known at NLO QCD~\cite{Campanario:2014b} and 
other contributions including notably tri-boson 
production processes with one boson decaying hadronicaly. 
The NLO QCD corrections to tri-boson 
production with leptonic decays were computed in
Refs.~\cite{Bozzi:2009ig,Bozzi:2010sj} and the hadronic decay modes are available via the
{\texttt{VBFNLO}} program~\cite{Arnold:2008rz,*Arnold:2011wj,*Baglio:2014uba}.
The interference effects between these contributions
are expected to be negligible for most measurements at the 
LHC~\cite{Campanario:2013gea}.

\begin{figure}[h]
  \centering
\includegraphics[width=0.45\textwidth]{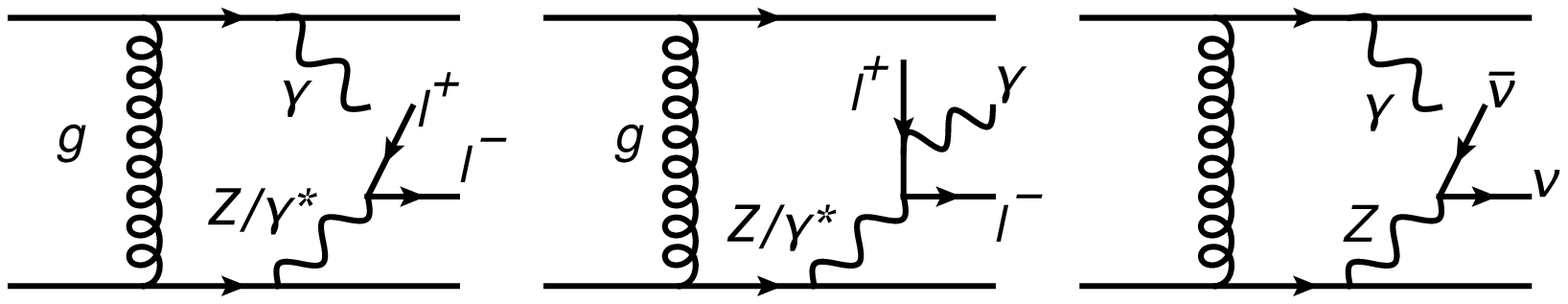}
\includegraphics[width=0.45\textwidth]{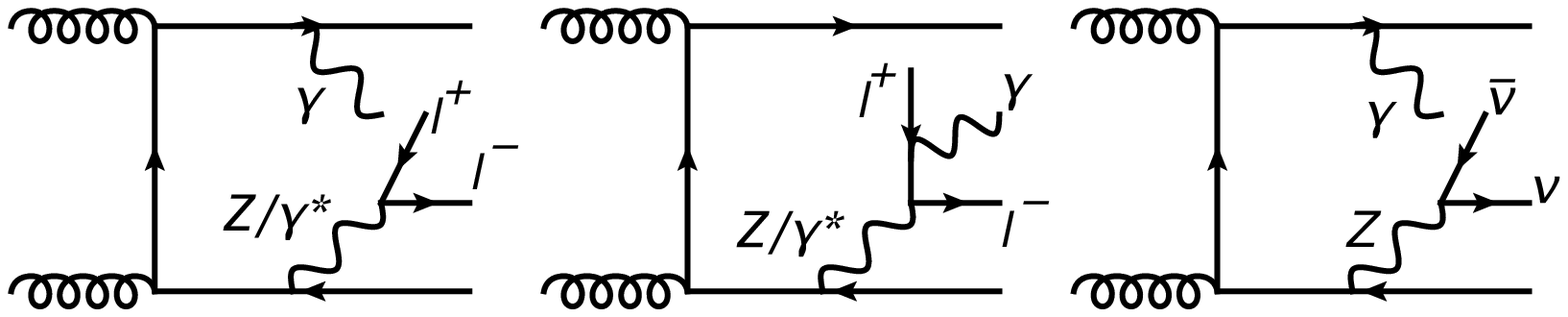}
\caption{Representative tree-level Feynman diagrams.}
\label{fig:feynTree}
\end{figure}

In this paper, we consider the QCD-induced mechanism for the processes
\begin{align}
pp &\to l^+ l^- \gamma j j + X, \qquad "Z_{l} \gamma jj" \label{processE}\\
pp &\to \bar{\nu}_l \nu_l\, \gamma j j\, + X,  \qquad "Z_{\nu} \gamma jj" \label{processN}
\end{align}
and will present the first theoretical prediction at NLO QCD accuracy~\footnote{Very recently, in Ref.~\cite{Alwall:2014hca}, results for the total cross section level for
on-shell $Z\gamma jj$ production have been reported.}.
Some representative Feynman diagrams at LO are displayed in \fig{fig:feynTree}. 
Since the dominant contribution comes from the
phase space region where the intermediate $Z$ boson is resonant, the above processes are
usually referred to as $Z_{l}\gamma jj$ and $Z_{\nu}\gamma jj$ 
production, accounting for the charged-lepton and neutrino pair production processes, respectively.
With this result, all the QCD-induced $VVjj$ production processes are
known at 
NLO QCD~\cite{Melia:2010bm,Melia:2011dw,Greiner:2012im,Denner:2012dz,Campanario:2013qba,Campanario:2013gea,Gehrmann:2013bga,Badger:2013ava,Campanario:2014dpa,Bern:2014vza,Campanario:2014ioa}.

The signature of an isolated photon together with two jets and missing energy is difficult to
study in experiment but is, as will be shown later, useful in a Monte Carlo analysis to find
(by comparing the two processes)
an optimal cut on the invariant mass of the two-charged lepton and photon system to remove the
radiative QED contribution (where the photon is emitted off the final charged leptons). This
contribution is unwanted because it reduces sensitivity to the weak boson scattering.
The focus of this paper is therefore on process~(\ref{processE}), however, a comparison of
normalized distributions to process~(\ref{processN}) will be performed.

We have implemented the QCD-induced processes (\ref{processE}) and (\ref{processN}) within the {\texttt{VBFNLO}}
framework~\cite{Arnold:2008rz,*Arnold:2011wj,*Baglio:2014uba}, a parton level Monte Carlo program
which allows the definition of general acceptance cuts and
distributions. As customary in {\texttt{VBFNLO}},
all off-shell effects, virtual photon contributions and spin-correlation effects
are fully taken into account. Our code will
be included in the next release of {\texttt{VBFNLO}}.

In the next section we sketch the calculational setup and
in Section~\ref{sec:results} we define our physical observables with a set of cuts and
present numerical results for the total cross section as well as various kinematical
distributions. Conclusions are presented in \sect{sec:concl}. 
Values of the virtual amplitudes at a random phase space point
are provided in the appendix to facilitate future
comparisons with our results.

\section{Calculational Setup}
\label{sec:calc}
The calculational method of the present paper follows closely the one presented in
\bib{Campanario:2014ioa} for the process $pp \to l_1^+ l_1^- l_2^+ l_2^- jj +X  $
~(called from now on $ZZjj$ for simplicity). 
As explained there, the gauge invariant class
of closed-quark loop diagrams with EW gauge bosons directly attached to the loop are discarded.
This contribution is at the few per mille level, hence negligible for all phenomenological purposes.
The diagrams with a closed-quark loop and two or three gluons attached to it are however included.
We work in the five-flavor scheme and virtual top loops are taken into account. We use the Frixione
isolation criteria~\cite{Frixione:1998jh} for the photon and 
therefore photon fragmentation functions are not included. 

Technically, the code for the $Z_{x} \gamma jj$ processes is adapted from the 
$ZZjj$ code with some modifications. This is 
possible because we use the effective current
approach and the spinor-helicity formalism~\cite{Hagiwara:1988pp,Campanario:2011cs}
factorizing the leptonic tensor containing the EW information of the system from the QCD amplitdue.
For the $l_1^+ l_1^- l_2^+ l_2^- jj$ case, the generic amplitudes for $V_1V_2jj$ with $V_i = Z/\gamma^*$ 
($i = 1,2$) and $\hat{V}jj$ with $\hat{V}=Z/\gamma^*$ are first created. Then the leptonic decays 
$V_i \to l_i^+ l_i^-$ and $\hat{V}\to l_1^+ l_1^- l_2^+ l_2^-$ are incorporated via 
effective currents. In this way, all off-shell effects and spin correlation are fully taken into account. 
This approach also makes it straightforward to obtain the $l^+ l^- \gamma j j$ 
and $\bar{\nu}_l \nu_l \gamma j j$ final states by picking the relevant generic amplitudes and 
changing the effective currents, namely, only the $Z\gamma jj$ generic amplitude and $Z\to \bar{\nu}_l \nu_l$ 
effective current are needed for the neutrino channel. For the charged-lepton case, 
we use the $V_1 \gamma jj$ and $\hat{V}^\prime jj$ generic amplitudes with $\hat{V}^\prime \to l^+ l^- \gamma$ 
effective current. 
These trivial changes are universal and have been crosschecked. 
Additionally, the phase-space generator has to be modified for a fast convergence of the Monte-Carlo integration. 
For this purpose, it is important to notice that, for on-shell photon production, there are two contributions dominating
in two different phase space regions associated with the two decay modes of the $Z$ bosons, namely
$Z \to l^+l^-$ and
$Z \to l^+l^-\gamma$.
This means that there are two different positions of the on-shell $Z$ pole in
the phase space for the process~(\ref{processE}). For efficient Monte
Carlo generation, we divide the phase space into two separate regions
to consider these two possibilities and then sum the two integrals to get the
total result. The regions are generated as double EW boson production
as well as $Z$ production with (approximately) on-shell $Z\to
l^+l^-\gamma$ three-body decay,
respectively, and
are chosen according to whether $m(l^+l^-\gamma)$ or $m(l^+l^-)$ is closer to $M_Z$.
The virtual photon contribution, which is far off-shell, does not pose additional 
problems and is always calculated together with the 
corresponding $Z$ contribution.
Another nontrivial change arises in the virtual amplitudes where we have to calculate a new set of scalar
integrals which do not occur in the off-shell photon case. We have again checked this with two independent
calculations (as explained in \bib{Campanario:2014ioa}) and obtained full agreement at the amplitude level.
Further details of our calculation and implementation and checks can be found in \bib{Campanario:2014ioa}. %
Furthemore, we have crosschecked the LO and real emission contributions without subtraction term against Sherpa~\cite{Gleisberg:2008ta,Gleisberg:2008fv} and agreement at the per mill level was found for
integrated cross sections. %

With this method, we obtain the NLO inclusive cross section with
statistical error of $1\%$ in 4 hours on an Intel $i7$-$3970X$
computer with one core and using the compiler Intel-ifort version
$12.1.0$. The distributions shown below are based on multiprocessor runs
with a total statistical error of  0.03\% at NLO. %
\section{Phenomenological results}
\label{sec:results}
For the numerical evaluation of the processes at the LHC operating at 14 TeV 
center-of-mass energy, we use the MSTW2008 parton distribution
function~\cite{Martin:2009iq} 
with $\alpha_s^{\LO(\NLO)}(M_Z)=0.13939 (0.12018)$ and the anti-$k_T$ 
cluster algorithm with a cone radius of
$R=0.4$. We consider jets with transverse momenta $p_{T,j}> 20 \GeV$ and rapidity
$|y_{j}|< 4.5$. To simulate experimental detector capabilities, we require
hard and central charged leptons with 
$p_{T,l}> 20 \GeV$ and
$|y_{l}|< 2.5$
and photons with $p_{T,\gamma}> 30 \GeV$ and
$|y_{\gamma}|< 2.5$. 
We impose minimal separation distances of $R_{jl}> 0.4$, $R_{ll} > 0.4$, $R_{l\gamma}> 0.4$ and $R_{j\gamma}> 0.7$. 
To avoid the need of including photon fragmentation functions, 
we use the photon isolation criteria \`a  la Frixione~\cite{Frixione:1998jh} 
with a cone radius of $\delta_0 = 0.7$.
Events are accepted if 
\bea
\sum_{i\in \text{partons}} p_{T,i}\theta(R-R_{\gamma i}) 
\le p_{T,\gamma}\frac{1-\cos R}{1-\cos\delta_0} \quad \forall R<\delta_0.
\label{eq:Frixione cut}
\eea
For the neutrinos of the ``$Z_{\nu} \gamma jj$ '' channel,
we do not apply any cut.
\begin{figure}[!h]
  \centering
\includegraphics[width=0.47\textwidth]{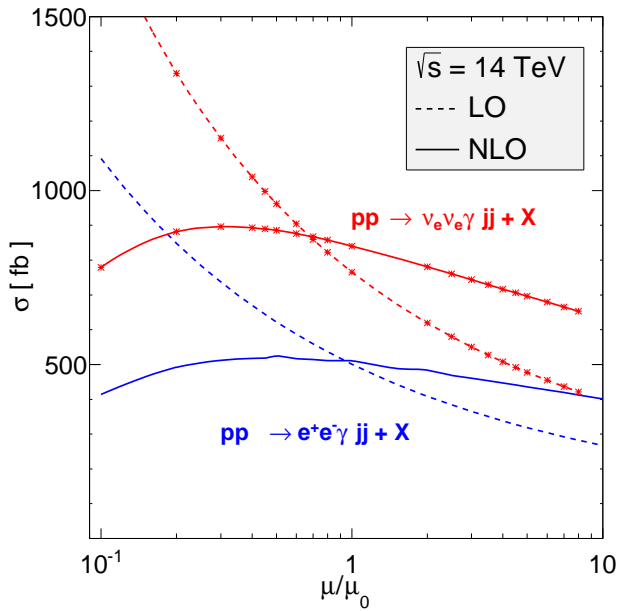}
  \caption{Scale dependence of the total LHC cross section at LO and NLO for
    $pp \to e^+ e^- \gamma jj + X $ and  $pp \to \nu_e \bar{\nu}_e \gamma jj+ X $ production around the central scale $\mu_0$ defined in
  Eq.~\ref{eq:scale}. The cuts used are described in the text.}
\label{fig:scale}
\end{figure}
\begin{figure*}[ht!]
  \centering
 \includegraphics[width=1\columnwidth]{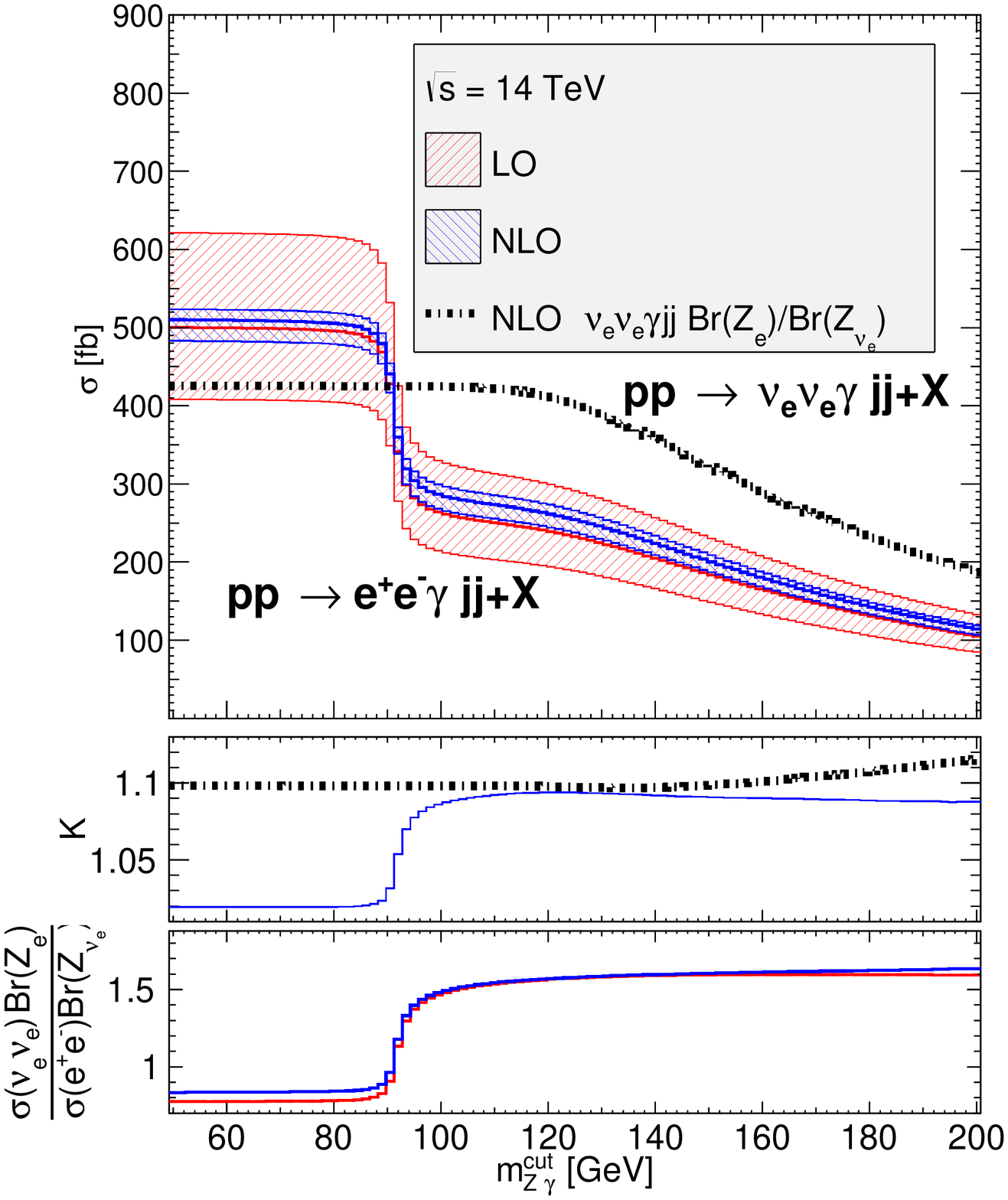} \hspace*{0.3cm}
  \includegraphics[width=1\columnwidth]{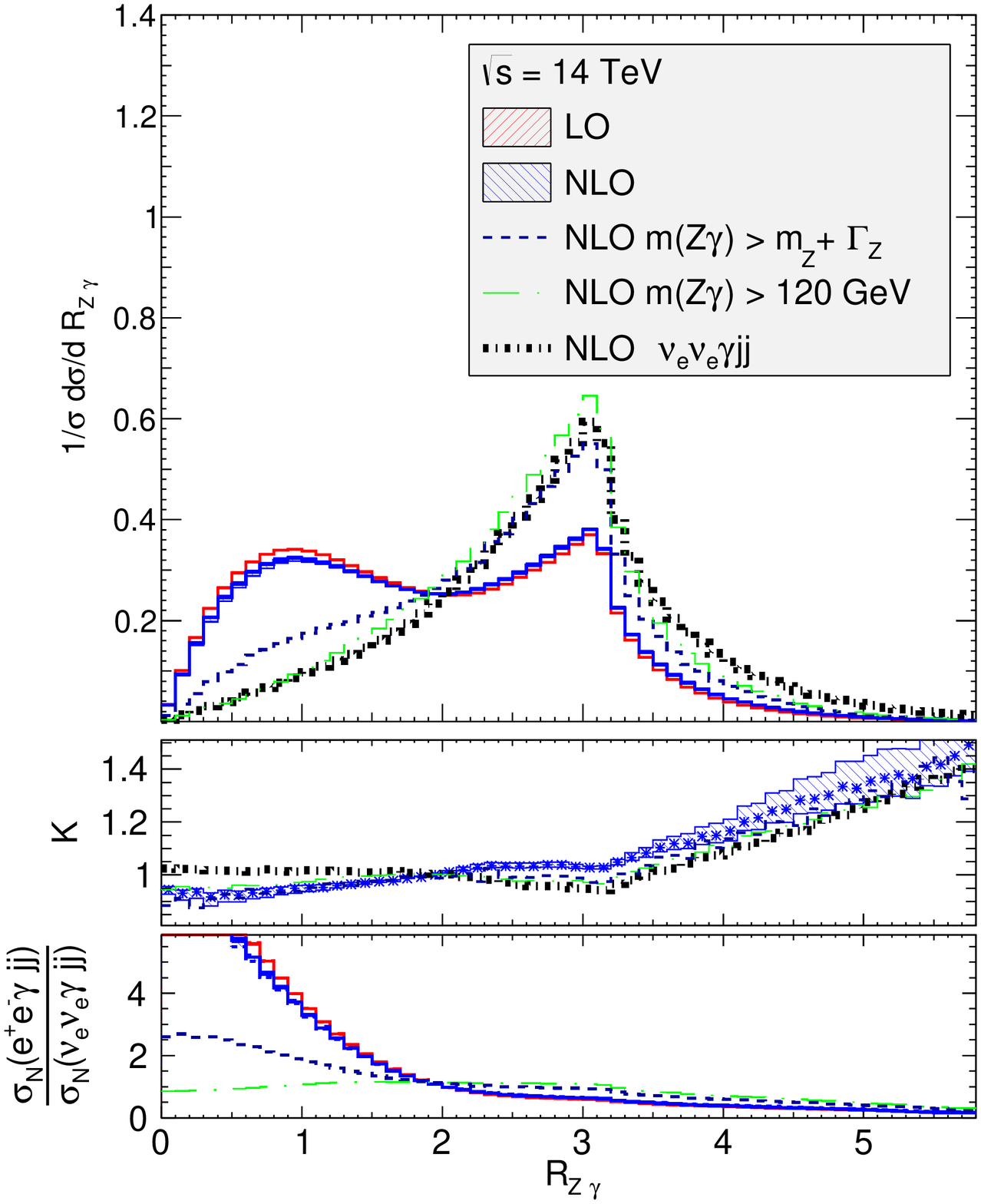}
   \caption{Left: Cross section for different values of the reconstructed
     $Z\gamma$ invariant mass cut. The neutrino curve is multiplied by the
     ratio of the charge-lepton versus neutrino branching ratios. The middle
     panel shows the K-factor and the lower the ratios of the modified neutrino
     cross section versus the LO and NLO electron cross sections. 
     Right: Normalized differential distributions of the rapidity-azimuthal angle separation 
     $R_{Z\gamma}$ for different values of the
     $m_{Z\gamma}^{\ensuremath{\,\mathrm{cut}}}$ cut. The middle and lower panels show the
     differential K-factor plots and the ratios of the normalized electron versus
     neutrino pair production channels.}
\label{fig:finalrad}
\end{figure*}
\begin{figure*}[ht!]
  \centering
  \includegraphics[width=1\columnwidth]{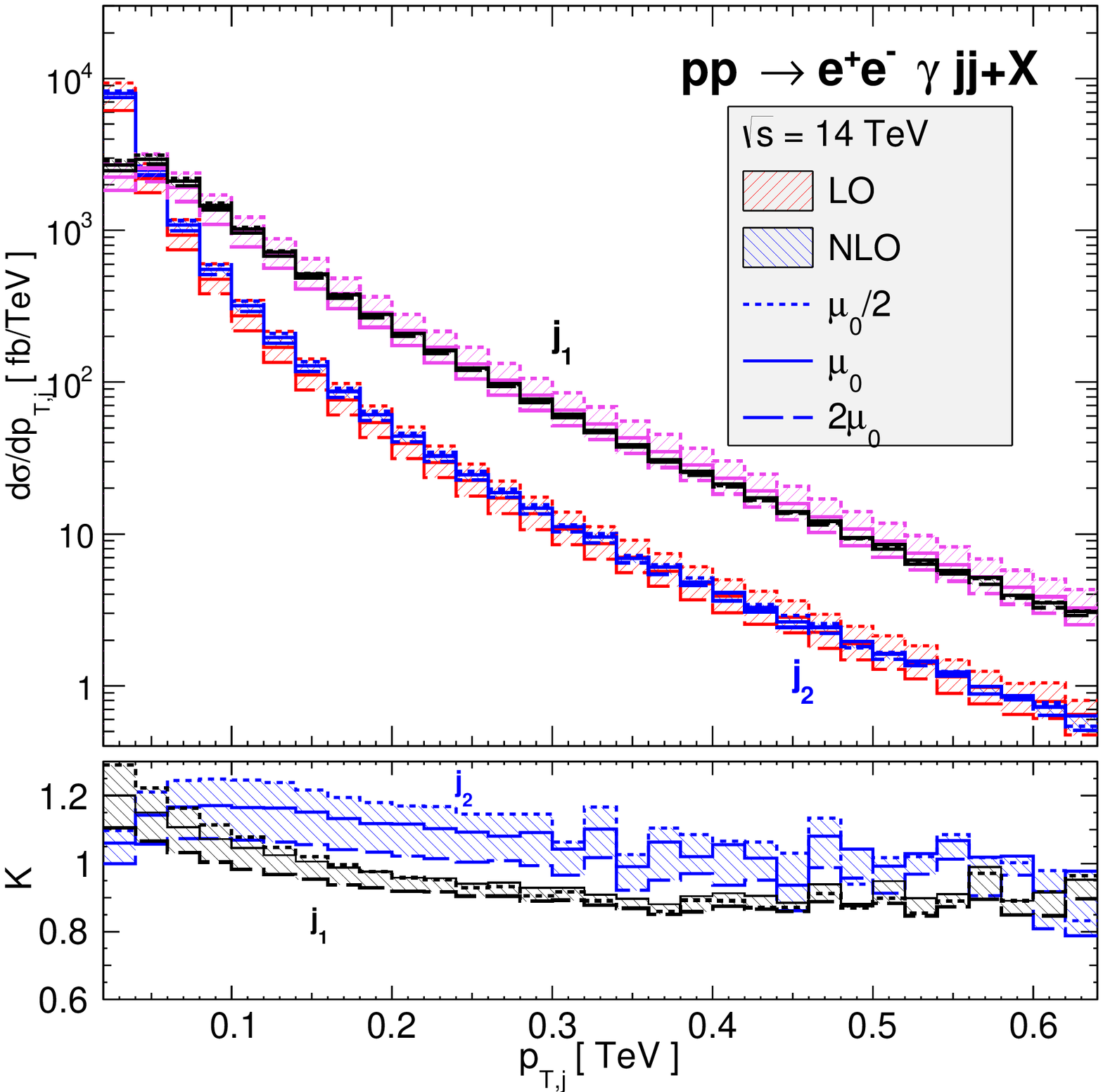} \hspace*{0.3cm}
  \includegraphics[width=1\columnwidth]{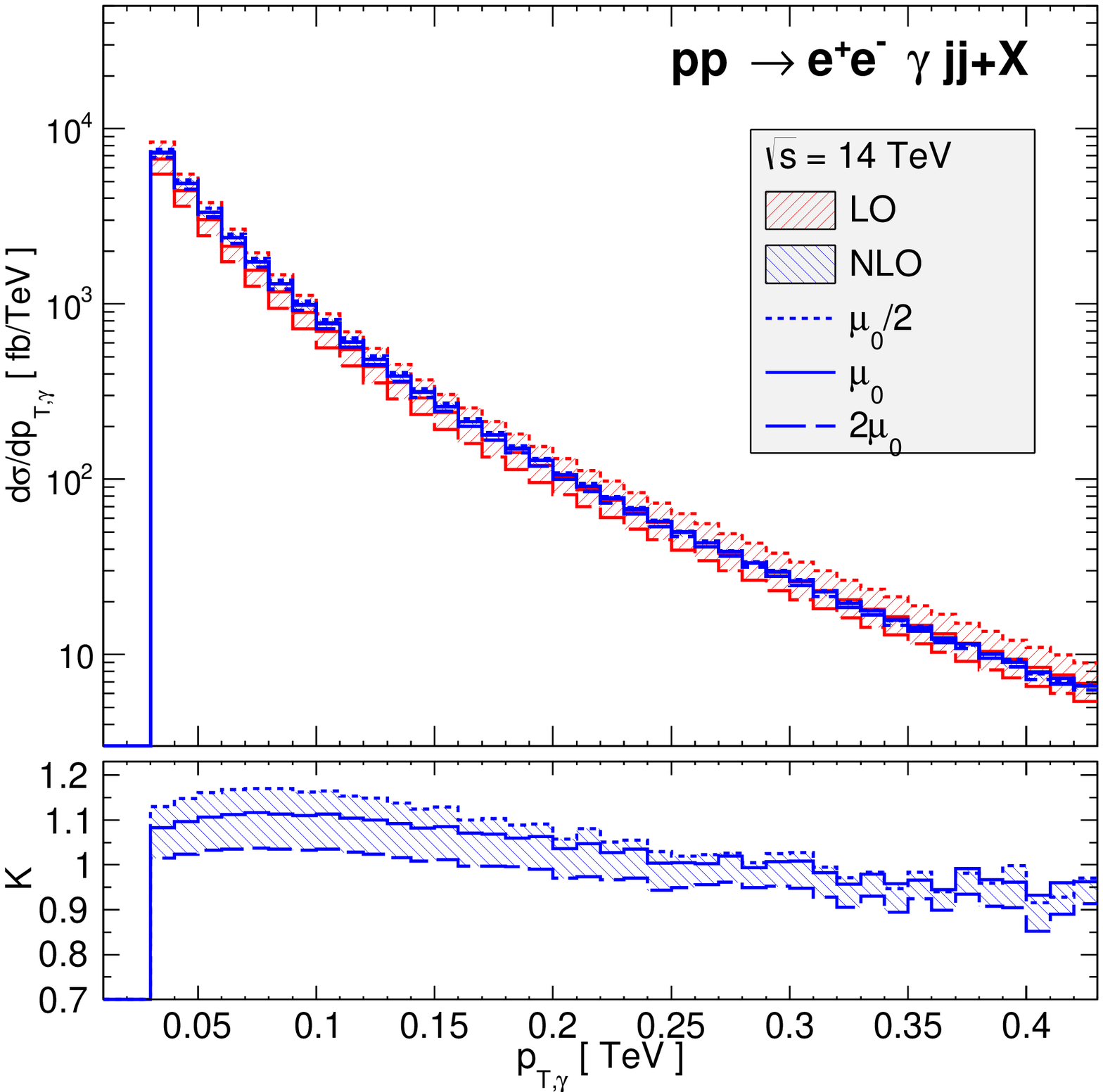}\\
  \includegraphics[width=1\columnwidth]{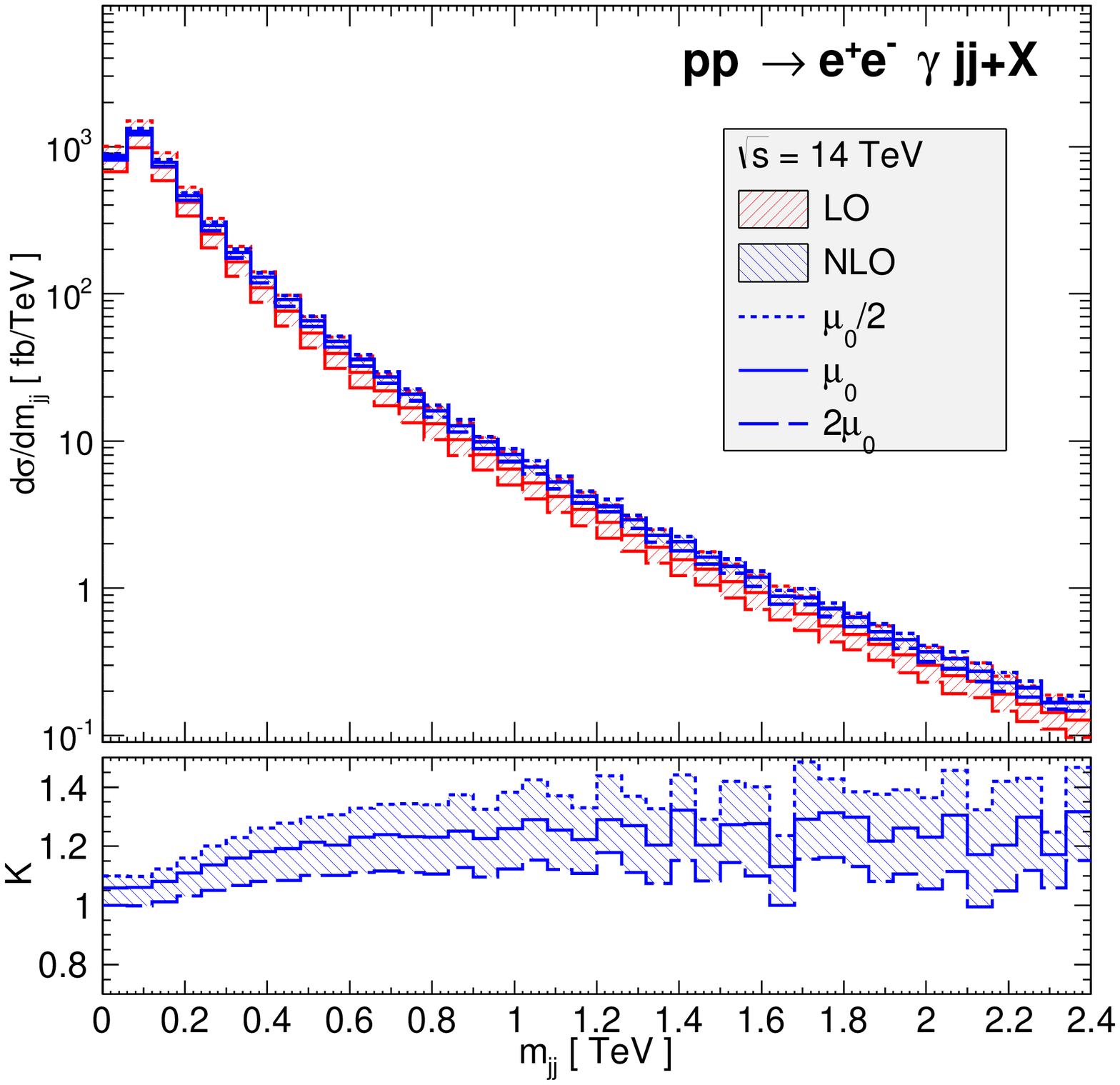} \hspace*{0.3cm}
  \includegraphics[width=1\columnwidth]{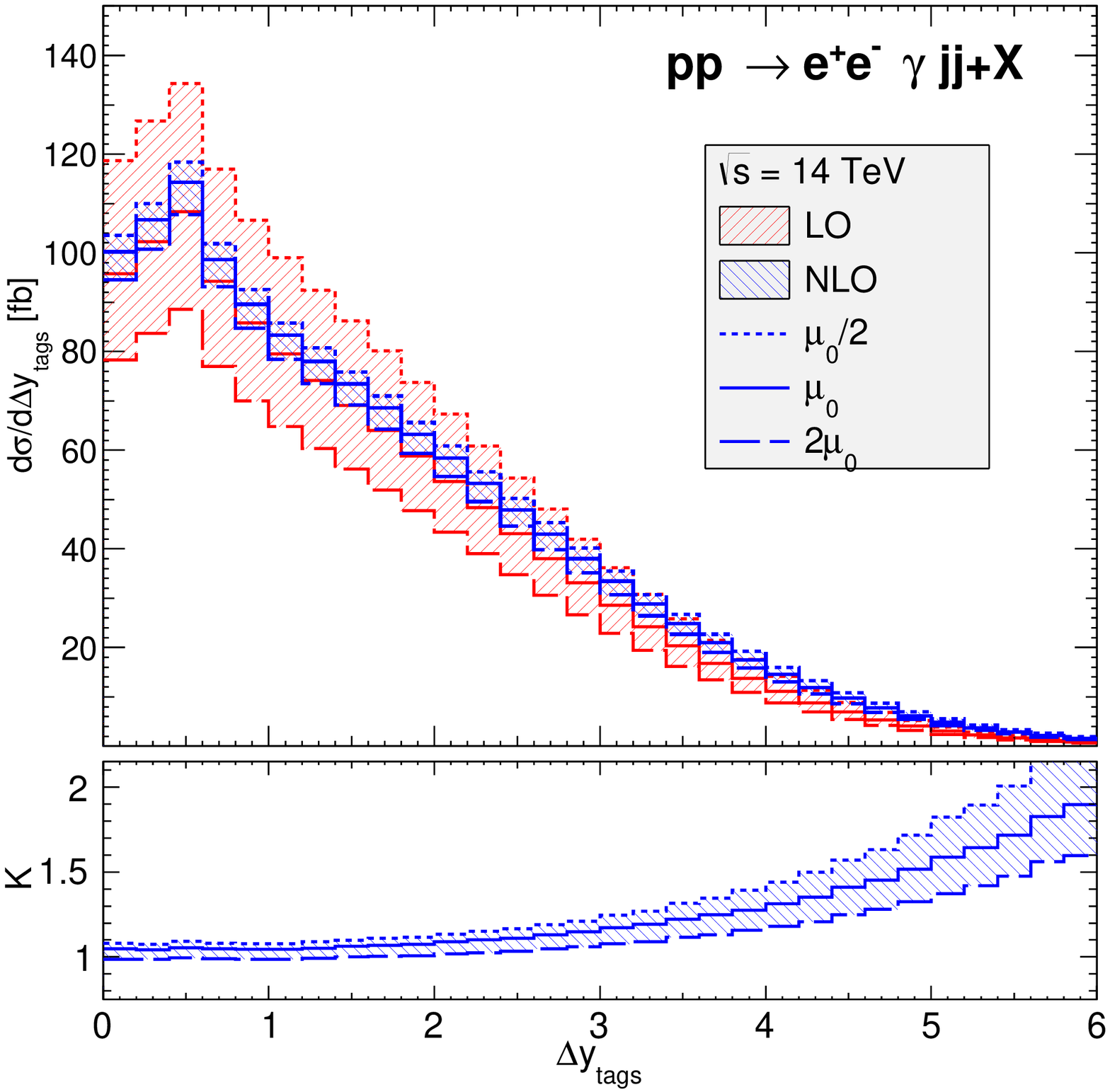}
  \caption{Differential distributions for the transverse momentum of the
    tagging jets (top left) and the photon (top right). On the bottom, the
    invariant mass (left) and rapidity separation (right) of the two tagging jets
    are displayed. The bands show the scale variations around the
    central scale, $\mu_0/2 \le \mu_F=\mu_R\le 2\mu_0$. In the small panels,
    the differential K-factors are plotted. The bands reflect the NLO scale
    variations with respect to $\sigma_\text{LO}(\mu_0)$. The inclusive cuts
    described in the text are used together with the cut on the invariant mass
    of the $Z\gamma$ system, $m_{Z\gamma}^{\text{cut}}=120 \GeV$, which
    eliminates the final radiative emission off the charged leptons.}
\label{fig:dist_NLO_jets_inc}
\end{figure*}

Other input parameters are chosen as $M_Z = 91.1876 \GeV$, $M_W=80.385 \GeV$ and
$G_F=1.16637 \times 10^{-5} \GeV^{-2}$. The electromagnetic 
coupling constant and the weak-mixing angle are calculated via tree level relations. 
All fermions are taken to be massless, except the top quark with $m_t=173.1 \GeV$. 
The width of the $Z$ is fixed at $\Gamma_Z= 2.508905 \GeV$. The strong coupling
constant is renormalized using the $\overline{\MS}$ scheme. The top-quark contribution 
is decoupled from the running, but is explicitly included in the one-loop amplitude. 
As a central factorization and renormalization scale, we use the sum of the transverse
energy $E_{T} = (p_T^2 + p^2)^{1/2}$ of the two tagging jets and of the
reconstructed $Z\gamma$ system,
\begin{equation}
\label{eq:scale}
\mu_{F} = \mu_{R} = \mu_{0}=\frac{1}{2} \left[E_{T}(jj) + E_{T}(VV)\right].
 \end{equation}
The first term interpolates between $m_{jj}$ and $\sum p_{T,jets}$ for
large and small $\Delta y_{jj}$ values, characterizing the dynamics of these processes appropriately.

In the following, we present results for the first generation of leptons. Taking into account both 
the electron and the muon yields an extra factor of two. Summing over three generations of neutrinos gives 
a factor of three. 

To evaluate the scale uncertainties associated to a fixed order calculation, we plot  in
Fig.~\ref{fig:scale} the cross section for the ``$Z_{\nu} \gamma jj$
'' and ``$Z_{l} \gamma jj$ '' channels varying the central scale in the range
$\mu \in (10^{-1},10) \mu_0$ simultaneously for the factorization and the
renormalization scale, which are set equal for simplicity. At the central
scale, we obtain $\sigma_{\LO}=500.82(3)^{+24\%}_{-18\%} \fb $ 
and $\sigma_{\NLO}= 510.6(1)^{+2.6\%}_{-5.3\%} \fb$ for the $e^+ e^- \gamma j j
$ channel and $\sigma_{\LO}=765.58(3)^{+26\%}_{-19\%}\fb$ and 
$\sigma_{\NLO}=840.8(3)^{+5.3\%}_{-7\%}\fb$ for the $\bar{\nu}_e \nu_e \gamma
j j$ one. The upper and lower numbers correspond to the scale uncertainties in
percentage for variations of a factor 2 around the central scale and the
number in brackets is the Monte-Carlo statistical error. At the central scale, we
observed very mild K-factors, defined as the ratio of the NLO over LO
predictions of the order of $1.01$ and $1.1$ for 
the ``$Z_{l} \gamma jj$'' and ``$Z_{\nu} \gamma jj$'' channels, respectively.

Next we investigate the radiative photon emission off the charged leptons 
in the ``$Z_{l} \gamma jj$ '' channel (see the middle Feynman diagrams of
Fig.~\ref{fig:feynTree}). These radiative decays present in both the EW and
QCD induced processes reduce the sensitivity to anomalous-coupling 
searches and therefore it is desirable to suppress them. 
This contribution dominates in the phase-space region where the reconstructed 
invariant mass of the $Z\gamma$ system is close to the $Z$ mass. 
Thus, imposing a cut on $M_{Z\gamma}$ around the $Z$ mass should remove this contribution. 
The optimum value of the cut is a priori uncertain. We therefore use 
the ``$Z_{\nu} \gamma jj$ '' channel, where radiative decays are absent, to determine it. 

In the left panel of Fig.~\ref{fig:finalrad}, as functions of the $m_{Z\gamma}$ cut, 
we plot the integrated NLO cross sections 
for the ``$Z_{l} \gamma jj$'' and  ``$Z_{\nu} \gamma jj$'' channels, the latter being multiplied by 
the ratio of the charge-lepton versus neutrino branching ratios of the $Z$, 
$Br(Z\to e^+ e^-)/Br(Z\to \bar{\nu}_e {\nu}_e) = 0.506$. 
The LO cross section is also shown for the ``$Z_{l} \gamma jj$'' channel. 
In the bottom panel, the ratios of the modified neutrino cross sections to the LO and NLO 
electron cross sections are plotted. 
They do not converge to one in the tails due to the different cuts applied for the charged leptons and the 
neutrinos. As expected, for the charged-lepton case, 
one observes that the cross section sharply decreases when the cut value is greater than 
the $Z$ mass. In the middle panel, the K-factors are plotted. We 
observe that $m_{Z\gamma}^{\text{cut}}=120 \GeV$ is a good value since 
the K-factor exhibits a plateau and the slope of the cross section curves are
approximately equal for both processes beyond this value (see bottom panel). This is confirmed in
the right panel of Fig.~\ref{fig:finalrad}, where the normalized differential
distributions of the reconstructed rapidity-azimuthal angle separation of the $Z\gamma$
system are plotted for the two channels. 
One observes that the cut $m_{Z\gamma} > m_Z+ \Gamma_Z$ reduces considerably 
the effect of the radiative decay in the charged-lepton channel, but some remnant 
is still clearly visible by comparing to the neutrino channel. Increasing the cut 
value to $120 \GeV$ makes the NLO distribution of the ``$Z_{l} \gamma jj$'' channel 
very similar to the corresponding ``$Z_{\nu} \gamma jj$'' one. 
This is better seen in the bottom panel, 
where the ratios of the normalized differential distributions between the two channels 
are plotted. The ratio of the $m_{Z\gamma}^{\text{cut}}=120 \GeV$ curve versus the ``$Z_{\nu}
\gamma jj$'' distribution is rather flat  and close to one till $R_{Z\gamma}$ reaches 
values of around 3 and then decreases. This difference is probably again 
due to the different cuts applied between the charged leptons and the neutrinos. 

In the following, we impose an additional cut $m_{Z\gamma}>120 \GeV$ and plot some 
relevant differential distributions for the two tagging jets and the photon 
at LO and NLO in the large panels of Fig.~\ref{fig:dist_NLO_jets_inc}. 
The tagging jets are defined as the two jets with highest transverse momenta and 
are ordered by hardness. 
The bands show the scale uncertainty in the range $\mu_0/2 \le \mu_F=\mu_R\le
 2\mu_0$. The small panels
always show the differential K-factors where the bands represents the scale variations of the NLO result, 
with respect to $\sigma_\text{LO}(\mu_0)$. In the top row, the
differential distributions of the transverse momentum of the two tagging jets (left) and
the photon (right) are plotted. The bottom row displays the invariant mass (left) and the
rapidity difference (right) of the two tagging jets. 
As expected, the scale uncertainty decreases considerably at NLO. Note that the
rapidity-separation distribution receives large NLO QCD corrections in the
region selected for vector boson fusion scattering, $\Delta y_\text{tags} >
3$. In general, the size of the K-factors range from $0.8$ to $1.9$, showing that NLO 
predictions are necessary for accurate measurements.

\section{Conclusions}
\label{sec:concl}
In this article, we have presented first results at NLO in QCD for the 
$pp \to l^+ l^- \gamma j j + X$ and $ pp \to \bar{\nu}_l \nu_l \gamma j j + X$
processes. With this result, all the QCD-induced $VVjj$ production processes are
known at NLO QCD. 

By comparing against the neutrino production process, we have been
able to efficiently remove the contribution of 
radiative photon emission off the charged leptons, which
diminishes the sensitivity of EW-induced processes to anomalous
couplings. As expected, the scale uncertainty is significantly reduced at NLO, 
which is visible both at the total and differential cross section level. 
The size of the NLO QCD corrections are phase space dependent ranging from $-20\%$
to $+90\%$, and are particularly large in the region where the vector-boson scattering signal is enhanced. 
NLO corrections are therefore needed for reliable predictions.

\section{Acknowledgments}
FC acknowledges financial support by the IEF-Marie Curie program (PIEF-GA-2011-298960) and partial funding by the  LHCPhenonet
(PITN-GA-2010-264564) and by the MINECO (FPA2011-23596).  MK is funded by the
graduate program GRK 1694: ``Elementary particle physics at highest  energy and
precision''. LDN and DZ are supported in part by the Deutsche Forschungsgemeinschaft
via the Sonderforschungsbereich/Transregio SFB/TR-9 ``Computational
Particle Physics''
%%%%%%%%%%%%%%%%%%%%%%%%%%%%%%%%%%%%%%%%%%%%%%%%%%%%%%%%%%%%%%

%\bibliographystyle{../h-physrev}
%\bibliography{../QCDVVjj}

\appendix*
\section{Results at one phase-space point}
\label{appendixA}
In the following, we present results at one phase-space point for the virtual amplitudes to
facilitate comparison with future calculations. 
We chose the same phase-space momentum configuration as \bib{Campanario:2014dpa}, \tab{table_PSP_2to5}, 
and give results for the process $j_1 j_2 \to j_3 j_4 e^+ e^-
\gamma$ at the squared-amplitude level, 
averaging over the initial-state helicities and colors. %
\begin{table*}[th]
 \begin{footnotesize}
 \begin{center}
 \caption{\label{table_PSP_2to5}{Momenta (in \GeV) at a random phase-space point for $j_1 j_2 \to j_3 j_4 e^+ e^- \gamma$ subprocesses.}}
%\vspace*{0.5cm}
\begin{tabular}{l | r@{.}l r@{.}l r@{.}l r@{.}l}
% \hline
& \multicolumn{2}{c}{ $E$}
& \multicolumn{2}{c}{ $p_x$}
& \multicolumn{2}{c}{ $p_y$}
& \multicolumn{2}{c}{ $p_z$}
\\
\hline
$j_1$  & 32&0772251055223 & 0&0  & 0&0  &  32&0772251055223  \\
$j_2$  & 2801&69305619768 & 0&0  & 0&0  &  -2801&69305619768  \\
$j_3$  & 226&525314156010 & -10&2177083492279 & -1&251308382450315$\times 10^{-15}$ & -226&294755550298 \\
$j_4$  & 327&281588297290 & -6&48554750244653 & -10&1061447270513 & -327&061219882068 \\
$e^+$  & 646&824307052136 & 36&0746355875450 & -26&0379256562231 & -645&292438579767 \\
$e^-$  & 1598&85193997112 & -2&88431497177613 & 24&4490976584709 & -1598&66239347157 \\
$\gamma$  & 34&2871318266438 & -16&4870647640944 & 11&6949727248035 & 27&6949763915464\\
\hline
\end{tabular}\end{center}
 \end{footnotesize}
\end{table*}
%%%
\begin{table*}[th]
 \begin{footnotesize}
 \begin{center}
\caption{\label{table_PSP_QCD_jjjj}{QCD interference amplitudes $2\text{Re}(\mathcal{A}_\text{NLO}\mathcal{A}^{*}_\text{LO})$
for $j_1 j_2 \to j_3 j_4 e^+ e^- \gamma$ subprocesses.}}
%\vspace*{0.5cm}
\begin{tabular}{l | r@{.}l r@{.}l r@{.}l}
\hline
& \multicolumn{2}{c}{ $1/\epsilon^2$}
& \multicolumn{2}{c}{ $1/\epsilon$}
& \multicolumn{2}{c}{ finite}
\\
\hline
& \multicolumn{2}{c}{ }
& \multicolumn{2}{c}{ $uu \to uu$}
& \multicolumn{2}{c}{ }
\\
\hline
I operator & 2&1930022552%51680
$\times 10^{-2}$ & 3&6933147142%30001
$\times 10^{-2}$ & 7&311094745$\times 10^{-2}$ \\
loop  & -2&19300225%5248082
$\times 10^{-2}$ & -3&6933147%14170613
$\times 10^{-2}$ & 0&14424709   \\
I+loop  & 3&6$\times 10^{-14}$ & 5&9$\times 10^{-13}$ & 0&2173580%37740763
\\
\hline
& \multicolumn{2}{c}{ }
& \multicolumn{2}{c}{ $uc \to uc$}
& \multicolumn{2}{c}{ }
\\
\hline
I operator & 3&1917977819%30914
$\times 10^{-2}$ & 5&1904760292%193195
$\times 10^{-2}$ & 0&1098476201 \\
loop  & -3&19179778%1921425
$\times 10^{-2}$  & -5&19047603%29082794
$\times 10^{-2}$ & 0&196364213%234776 
\\
I+loop  & 9&5$\times 10^{-14}$ & 1&1$\times 10^{-12}$ & 0&30621183%3364361
 \\\hline
& \multicolumn{2}{c}{ }
& \multicolumn{2}{c}{ $ud \to ud$}
& \multicolumn{2}{c}{ }
\\
\hline
I operator & 5&9210009570%68433
$\times 10^{-2}$ & 9&6286844080%39630
$\times 10^{-2}$ & 0&2037747651%33858 
\\
loop  & -5&92100095%7038619
$\times 10^{-2}$  & -9&628684%263502846
$\times 10^{-2}$ & 0&53005178%4374044 
\\
I+loop  & 3&0$\times 10^{-13}$ & 1&4$\times 10^{-9}$ & 0&7338265%49507902
\\
\hline
& \multicolumn{2}{c}{ }
& \multicolumn{2}{c}{ $dd \to dd$}
& \multicolumn{2}{c}{ }
\\
\hline
I operator & 6&8465222944%50105
$\times 10^{-3}$ & 1&1487721719%98274
$\times 10^{-2}$ & 2&246881141%082626
$\times 10^{-2}$ \\
loop  & -6&84652229%4425448
$\times 10^{-3}$  & -1&14877217%1872908
$\times 10^{-2}$ & 4&789599960%288074
$\times 10^{-2}$ \\
I+loop  & 2&5$\times 10^{-14}$ & 2&5$\times 10^{-13}$ & 7&03648110%1370701
$\times 10^{-2}$ \\
\hline
& \multicolumn{2}{c}{ }
& \multicolumn{2}{c}{ $ds \to ds$}
& \multicolumn{2}{c}{ }
\\
\hline
I operator & 1&04197858244%5794
$\times 10^{-2}$ & 1&6944572384%04918
$\times 10^{-2}$ & 3&586031267%069384
$\times 10^{-2}$ \\
loop  & -1&041978582%440196
$\times 10^{-2}$  & -1&69445724%38356271
$\times 10^{-2}$ & 6&930377376%103564
$\times 10^{-2}$ \\
I+loop  & 5&6$\times 10^{-14}$ & 4&9$\times 10^{-13}$ & 0&105164086%431729
\\
\hline
& \multicolumn{2}{c}{ }
& \multicolumn{2}{c}{ $gg \to \bar{u}u$}
& \multicolumn{2}{c}{ }
\\
\hline
I operator & 4&2224900184%74755
$\times 10^{-4}$ & 2&5147277683%50466
$\times 10^{-5}$ & 5&73018157%2404789
$\times 10^{-4}$ \\
loop  & -4&2224900%18213713
$\times 10^{-4}$  & -2&51472%7743965480
$\times 10^{-5}$ & 1&24085333%0835127
$\times 10^{-3}$ \\
I+loop  & 2&6$\times 10^{-14}$ & 2&4$\times 10^{-13}$ & 1&81387148%8075606
$\times 10^{-3}$ \\
\hline
& \multicolumn{2}{c}{ }
& \multicolumn{2}{c}{ $gg \to \bar{d}d$}
& \multicolumn{2}{c}{ }
\\
\hline
I operator & 1&13509295313%4539
$\times 10^{-4}$ & 6&3611573393%18822
$\times 10^{-6}$ & 1&56909560%3958936
$\times 10^{-4}$ \\
loop  & -1&135092953%129129
$\times 10^{-4}$  & -6&36116%57314958587
$\times 10^{-6}$ & 3&27276628%79617286
$\times 10^{-4}$ \\
I+loop  & 5&4$\times 10^{-16}$ & 2&4$\times 10^{-14}$ & 4&8418618%83576221
$\times 10^{-4}$ \\
\hline
\end{tabular}\end{center}
 \end{footnotesize}
\end{table*}
We include all UV counterterms and all closed-quark loops with
gluons attached to it. Diagrams including a closed-quark loop with the
$Z/\gamma^*$ directly attached to it are excluded. 
The top quark is decoupled from the running of $\alpha_s$. 
However, its contribution is explicitly included in the one-loop amplitudes. 
Here we use $\alpha = \alpha_s = 1$ for simplicity. 
With this set up and measuring energies in \GeV, we get at tree level,
\begin{align}
  \overline{|\mathcal{A}_\text{LO}^{uu\rightarrow uu}|}^2 &= 2.583569915405990\times 10^{-2},\nonumber \\
  \overline{|\mathcal{A}_\text{LO}^{uc\rightarrow uc}|}^2 &= 3.760248173799574\times 10^{-2},\nonumber \\
  \overline{|\mathcal{A}_\text{LO}^{ud\rightarrow ud}|}^2 &=
  6.975514915738625\times 10^{-2},\nonumber 
\end{align} 
\begin{align}
  \overline{|\mathcal{A}_\text{LO}^{dd\rightarrow dd}|}^2 &= 8.065869053590906\times 10^{-3},\nonumber \\
  \overline{|\mathcal{A}_\text{LO}^{ds\rightarrow ds}|}^2 &= 1.227552097429276\times 10^{-2},\nonumber \\
  \overline{|\mathcal{A}_\text{LO}^{gg\rightarrow \bar{u}u}|}^2 &= 3.061233143517198\times 10^{-4},\nonumber \\
  \overline{|\mathcal{A}_\text{LO}^{gg\rightarrow \bar{d}d}|}^2 &= 8.229230037027499\times 10^{-5}.
\end{align} 
For the one-loop integrals, we use the convention
\bea
T_0 = \frac{\mu_R^{2\epsilon}\Gamma(1-\epsilon)}{i\pi^{2-\epsilon}}\int d^D q \frac{1}{(q^2 - m_1^2 + i0)\cdots},
\eea
with $D=4-2\epsilon$. Additionally, the conventional dimensional regularization method~\cite{'tHooft:1972fi}
with $\mu_{R} = M_Z$ is used.
With this, the interference amplitudes
$2\text{Re}(\mathcal{A}_\text{NLO}\mathcal{A}^{*}_\text{LO})$,
for the one-loop corrections
and the I-operator contribution as defined in \bib{Catani:1996vz},
are given in \tab{table_PSP_QCD_jjjj}. 

Switching from the conventional dimensional regularization to dimensional reduction method
induces a finite shift, which can be calculated noting that
the sum $|\mathcal{A}_\text{LO}|^2+2\text{Re}(\mathcal{A}_\text{NLO}\mathcal{A}^{*}_\text{LO})$
should be constant~\cite{Catani:1996pk}.
The shift
on the Born amplitude squared comes from the change in the strong coupling constant,
see e.g. \bib{Kunszt:1993sd},
\bea
\alpha_s^{\overline{DR}} = \alpha_s^{\overline{MS}}\left(1+\frac{\alpha_s}{4\pi}\right).
\eea
Finally, using the rule given in \bib{Catani:1996vz}, the shift on the I-operator contribution can be calculated.

%%%%%%%%%%%%%%%%%%%%%%%%%%%%%%%%%%%%%%%%%%%%%%%%%%%%%%%%%%%%%%
\bibliographystyle{h-physrev}
\bibliography{QCDVVjj}
\end{document}